\documentclass[12pt]{article}
\usepackage{times}
\usepackage{geometry}
\usepackage[utf8]{inputenc}
\usepackage{enumitem,amssymb}
\usepackage{ragged2e}
\newlist{thematic}{itemize}{8}
\setlist[thematic]{label=$\square$}
\usepackage{pifont}
\usepackage{graphicx}				
\usepackage{amssymb,amsmath}
\usepackage[multiple]{footmisc}
\usepackage{aas_macros}
\usepackage{hyperref}
\usepackage{titlesec}

\titleformat*{\section}{\large\bfseries}

\begin{document}
\begin{raggedright}
\huge
Astro2020 Science White Paper \linebreak
Testing Gravity Using Type Ia Supernovae Discovered by Next-Generation Wide-Field Imaging Surveys \linebreak
\normalsize

\noindent \textbf{Thematic Areas:} \hspace*{60pt} $\square$ Planetary Systems \hspace*{10pt} $\square$ Star and Planet Formation \hspace*{20pt}\linebreak
$\square$ Formation and Evolution of Compact Objects \hspace*{31pt} $\blacksquare$ Cosmology and Fundamental Physics \linebreak
  $\square$  Stars and Stellar Evolution \hspace*{1pt} $\square$ Resolved Stellar Populations and their Environments \hspace*{40pt} \linebreak
  $\square$    Galaxy Evolution   \hspace*{45pt} $\square$             Multi-Messenger Astronomy and Astrophysics \hspace*{65pt} \linebreak
  
\textbf{Principal Author:}

Name:	
 Alex Kim\linebreak						
Institution:  
Physics Division, Lawrence Berkeley National Laboratory, 
    1 Cyclotron Road, Berkeley, CA, 94720 \linebreak
Email:  \url{agkim@lbl.gov}
 \linebreak
Phone:  1-510-486-4621
 \linebreak
 
\textbf{Co-authors:}
G.~Aldering\footnotemark[1],
\footnotetext[1]{Physics Division, Lawrence Berkeley National Laboratory, 
    1 Cyclotron Road, Berkeley, CA, 94720}
P.~Antilogus\footnotemark[2],
\footnotetext[2]{Laboratoire de Physique Nucl\'eaire et de Hautes Energies, Sorbonne Universit\'e, CNRS-IN2P3, 4 Place Jussieu, 75005 Paris, France}    
A.~Bahmanyar\footnotemark[3],
\footnotetext[3]{Department of Astronomy and Astrophysics, University of Toronto,
50 St. George Street, Toronto, Ontario, Canada M5S 3H4}
S.~BenZvi\footnotemark[4],
\footnotetext[4]{Department of Physics and Astronomy, University of Rochester, Rochester, NY 14627, USA}
H.~Courtois\footnotemark[5],
\footnotetext[5]{Universit\'e de Lyon, F-69622, Lyon, France; Universit\'e de Lyon
1, Villeurbanne; CNRS/IN2P3, Institut de Physique Nucl\'eaire de
Lyon, France} 
T.~Davis\footnotemark[6],
\footnotetext[6]{School of Mathematics and Physics, University of Queensland, Brisbane, QLD 4072, Australia}
H.~Feldman\footnotemark[7],
\footnotetext[7]{Department of Physics \& Astronomy, University of Kansas, Lawrence, KS 66045-7550 USA}
S.~Ferraro\footnotemark[1],
S.~Gontcho A Gontcho\footnotemark[4],
O.~Graur\footnotemark[8]\textsuperscript{,}\footnotemark[9]\textsuperscript{,}\footnotemark[10],
\footnotetext[8]{Harvard-Smithsonian Center for Astrophysics, 60 Garden Street, Cambridge, MA 02138, USA}
\footnotetext[9]{Department of Astrophysics, American Museum of Natural History, New York, NY 10024, USA}
\footnotetext[10]{NSF Astronomy and Astrophysics Postdoctoral Fellow}
R.~Graziani\footnotemark[11],
\footnotetext[11]{Universit\'e Clermont Auvergne, CNRS/IN2P3, Laboratoire de
Physique de Clermont, F-63000 Clermont-Ferrand, France}
J.~Guy\footnotemark[1],
C.~Harper\footnotemark[1],
R.~Hlo\v{z}ek\footnotemark[3]\textsuperscript{,}\footnotemark[12],
\footnotetext[12]{Dunlap Institute for Astronomy and Astrophysics, University of Toronto, ON, M5S 3H4}
C.~Howlett\footnotemark[13],
\footnotetext[13]{International Centre for Radio Astronomy Research, The University of Western Australia, Crawley, WA 6009, Australia}
D.~Huterer\footnotemark[14],
\footnotetext[14]{Department of Physics, University of Michigan, 450 Church Street, Ann
Arbor, MI 48109, USA }
C.~Ju\footnotemark[1],
P.-F.~Leget\footnotemark[15],
\footnotetext[15]{Laboratoire de Physique Nucl\'eaire et de Hautes Energies, Sorbonne Universit\'e, CNRS-IN2P3, 4 Place Jussieu, 75005 Paris, France}    
E.~V.~Linder\footnotemark[1],
P.~McDonald\footnotemark[1],
J.~Nordin\footnotemark[16],
\footnotetext[16]{Institut fur Physik, Humboldt-Universitat zu Berlin, Newtonstr. 15, 12489 Berlin, Germany}
P.~Nugent\footnotemark[17],
\footnotetext[17]{Computational Research Division, Lawrence Berkeley National Laboratory, 
    1 Cyclotron Road, Berkeley, CA, 94720}
S.~Perlmutter\footnotemark[1]\textsuperscript{,}\footnotemark[18],
\footnotetext[18]{Department of Physics, University of California Berkeley, Berkeley, CA 94720}
N.~Regnault\footnotemark[15],
M.~Rigault\footnotemark[11],
A.~Shafieloo\footnotemark[19],
\footnotetext[19]{Korea Astronomy and Space Science Institute, Yuseong-gu, Daedeok-daero 776, Daejeon 34055, Korea}
A.~Slosar\footnotemark[20],
\footnotetext[20]{Brookhaven National Laboratory, Physics Department, Upton, NY
11973, USA} 
R.~B.~Tully\footnotemark[21],
\footnotetext[21]{Institute for Astronomy, University of Hawaii, 2680 Woodlawn Drive, Honolulu, HI 96822, USA}
L.~Wang\footnotemark[22],
\footnotetext[22]{Texas A\&M University, College Station, TX 77843}
M.~White\footnotemark[1]\textsuperscript{,}\footnotemark[18],
M.~Wood-Vasey\footnotemark[23]
\footnotetext[23]{PITT PACC, Department of Physics and Astronomy, University of Pittsburgh, Pittsburgh, PA 15260, USA} 
\linebreak

\textbf{Endorsers:}
Behzad Ansarinejad
\footnotemark[24],
\footnotetext[24]{Department of Physics, Lower Mountjoy, South Rd, Durham DH1 3LE, United Kingdom} 
Robert Armstrong\footnotemark[25],
\footnotetext[25]{Lawrence Livermore National Laboratory, Livermore, CA, 94550} 
Jacobo Asorey\footnotemark[19],
Carlo Baccigalupi\footnotemark[26]\textsuperscript{,}\footnotemark[27]\textsuperscript{,}\footnotemark[28],
\footnotetext[26]{SISSA - International School for Advanced Studies, Via Bonomea 265, 34136 Trieste, Italy} 
\footnotetext[27]{IFPU - Institute for Fundamental Physics of the Universe, Via Beirut 2, 34014 Trieste, Italy} 
\footnotetext[28]{INFN - National Institute for Nuclear Physics, Via Valerio 2, I-34127 Trieste, Italy} 
Maciej Bilicki\footnotemark[29],
\footnotetext[29]{Center for Theoretical Physics, Polish Academy of Sciences, al. Lotnik\'{o}w 32/46, 02-668, Warsaw, Poland} 
Julian Borrill\footnotemark[17], 
Elizabeth Buckley-Geer\footnotemark[30],
\footnotetext[30]{Fermi National Accelerator Laboratory, Batavia, IL 60510} 
Kelly A. Douglass\footnotemark[4],
Cora Dvorkin\footnotemark[31],
\footnotetext[31]{Department of Physics, Harvard University, Cambridge, MA 02138, USA} 
Simon Foreman\footnotemark[32],
\footnotetext[32]{Canadian Institute for Theoretical Astrophysics, University of Toronto, Toronto, ON M5S 3H8, Canada} 
Llu\'is Galbany\footnotemark[23],
Juan Garc\'ia-Bellido\footnotemark[33]\textsuperscript{,}\footnotemark[34],
\footnotetext[33]{Instituto de Fisica Teorica UAM/CSIC, Universidad Autonoma de Madrid, 28049 Madrid, Spain} 
\footnotetext[34]{Universidad Aut\'{o}noma de Madrid, 28049, Madrid, Spain} 
Martina Gerbino\footnotemark[35],
\footnotetext[35]{HEP Division, Argonne National Laboratory, Lemont, IL 60439, USA} 
Mandeep S.S. Gill\footnotemark[36]\textsuperscript{,}\footnotemark[37]\textsuperscript{,}\footnotemark[38],
\footnotetext[36]{Kavli Institute for Particle Astrophysics and Cosmology, Stanford 94305} 
\footnotetext[37]{Stanford University, Stanford, CA 94305} 
\footnotetext[38]{SLAC National Accelerator Laboratory, Menlo Park, CA 94025} 
Larry Gladney\footnotemark[39],
\footnotetext[39]{Department of Physics, Yale University, New Haven, CT 06520} 
Saurabh W.~Jha\footnotemark[40],
\footnotetext[40]{Department of Physics and Astronomy, Rutgers, the State University of New Jersey, 136 Frelinghuysen Road, Piscataway, NJ 08854, USA} 
Johann Cohen-Tanugi\footnotemark[41],
\footnotetext[41]{Laboratoire Univers et Particules de Montpellier, Univ. Montpellier and CNRS, 34090 Montpellier, France} 
D.\ O.\ Jones\footnotemark[42],
\footnotetext[42]{University of California at Santa Cruz, Santa Cruz, CA 95064} 
Marc Kamionkowski\footnotemark[43],
\footnotetext[43]{Johns Hopkins University, Baltimore, MD 21218} 
Ryan E. Keeley\footnotemark[19],
Robert Kehoe\footnotemark[44],
\footnotetext[44]{Southern Methodist University, Dallas, TX 75275} 
Savvas M. Koushiappas\footnotemark[45],
\footnotetext[45]{Brown University, Providence, RI 02912} 
Ely D.~Kovetz\footnotemark[46],
\footnotetext[46]{Department of Physics, Ben-Gurion University, Be'er Sheva 84105, Israel} 
Kazuya Koyama\footnotemark[47],
\footnotetext[47]{Institute of Cosmology \& Gravitation, University of Portsmouth, Dennis Sciama Building, Burnaby Road, Portsmouth PO1 3FX, UK} 
Benjamin L'Huillier\footnotemark[19],
Ofer Lahav\footnotemark[48],
\footnotetext[48]{University College London, WC1E 6BT London, United Kingdom} 
Chien-Hsiu Lee\footnotemark[49],
\footnotetext[49]{National Optical Astronomy Observatory, 950 N. Cherry Ave., Tucson, AZ 85719 USA} 
Michele Liguori\footnotemark[50],
\footnotetext[50]{Dipartimento di Fisica e Astronomia ``G. Galilei'',Universit\`a degli Studi di Padova, via Marzolo 8, I-35131, Padova, Italy} 
Axel de la Macorra\footnotemark[51],
\footnotetext[51]{IFUNAM - Instituto de F\'{i}sica, Universidad Nacional Aut\'onoma de M\'exico, 04510 CDMX, M\'exico} 
Joel Meyers\footnotemark[44],
Surhud More\footnotemark[52],
\footnotetext[52]{The Inter-University Centre for Astronomy and Astrophysics, Pune, 411007, India} 
Jeffrey A. Newman\footnotemark[23],
Gustavo Niz\footnotemark[53],
\footnotetext[53]{Divisi\'{o}n de Ciencias e Ingenier\'{i}as, Universidad de Guanajuato, Le\'{o}n 37150, M\'{e}xico} 
Antonella Palmese\footnotemark[30],
Francesco Piacentini\footnotemark[54]\textsuperscript{,}\footnotemark[55],
\footnotetext[54]{Dipartimento di Fisica, Universit\`{a} La Sapienza, P. le A. Moro 2, Roma, Italy} 
\footnotetext[55]{Istituto Nazionale di Fisica Nucleare, Sezione di Roma, 00185 Roma, Italy} 
Steven Rodney\footnotemark[56],
\footnotetext[56]{The University of South Carolina, Columbia, SC 29208} 
Benjamin Rose\footnotemark[57],
\footnotetext[57]{Space Telescope Science Institute, Baltimore, MD 21218} 
Masao Sako\footnotemark[58],
\footnotetext[58]{Department of Physics and Astronomy, University of Pennsylvania, Philadelphia, Pennsylvania 19104, USA} 
Lado Samushia\footnotemark[59],
\footnotetext[59]{Kansas State University, Manhattan, KS 66506} 
Neelima Sehgal\footnotemark[60],
\footnotetext[60]{Stony Brook University, Stony Brook, NY 11794} 
Sara Simon\footnotemark[14],
Stephanie Escoffier\footnotemark[61],
\footnotetext[61]{Aix Marseille Univ, CNRS/IN2P3, CPPM, Marseille, France} 
Scott Watson\footnotemark[62],
\footnotetext[62]{Syracuse University, Syracuse, NY 13244} 
Weishuang Xu\footnotemark[31],
Gong-Bo Zhao\footnotemark[63]\textsuperscript{,}\footnotemark[47],
\footnotetext[63]{National Astronomical Observatories, Chinese Academy of Sciences, PR China} 
Yi Zheng\footnotemark[64]
\footnotetext[64]{School of Physics, Korea Institute for Advanced Study, 85 Hoegiro, Dongdaemun-gu, Seoul 130-722, Korea} 

\textbf{Abstract:}
In the upcoming decade cadenced wide-field imaging surveys 
will increase the number of identified  $z<0.3$ Type~Ia supernovae (SNe~Ia)  from the hundreds to the
hundreds of thousands.  The increase in the number density  and solid-angle coverage 
of SNe~Ia, in parallel with improvements in the standardization of
their absolute magnitudes, now make them competitive probes of the growth of structure and hence of gravity.  The peculiar velocity power spectrum
is sensitive to the growth index $\gamma$, which captures the effect of gravity on the linear growth of structure through the relation $f=\Omega_M^\gamma$.
We present the first  projections for the precision in $\gamma$ for a range of realistic SN peculiar-velocity survey scenarios.
In the next decade the peculiar velocities of
SNe~Ia in the local $z<0.3$ Universe will provide a measure of $\gamma$ to $\pm 0.01$ precision that can definitively distinguish  between General Relativity and leading models of alternative gravity.
\end{raggedright}
\pagebreak
\newpage
\setcounter{page}{0}
\pagenumbering{arabic}
\setcounter{page}{1}

\section{Introduction}
In the late 1990's, Type~Ia supernovae (SNe~Ia) were used as distance probes to measure the homogeneous expansion history of the Universe.  The remarkable discovery
that the expansion is accelerating  has called into question our basic understanding of the gravitational forces within the Universe.  Either it
is dominated by a ``dark energy'' that is gravitationally repulsive, or General Relativity is inadequate and needs to be replaced by a modified theory of
gravity.  It is only appropriate that in the upcoming decade, with their sheer numbers, solid-angle coverage,
and improved distance precisions, SNe~Ia will provide measurements of the {\it inhomogeneous} motions of structures in the Universe
that will provide an unmatched test of whether dark energy or modified gravity is responsible for the accelerating expansion of the Universe.

In the next decade, SNe~Ia will be used as peculiar-velocity probes to measure  the influence of gravity on structure formation within the Universe.
Peculiar velocities induce scatter along the redshift axis of the SN Hubble diagram, which is
pronounced at low redshifts and when the magnitude scatter (e.g.\ due to intrinsic magnitude dispersion) is small.
The peculiar velocity power spectrum is sensitive to the growth of structure as $P_{vv}\propto (fD)^2$, where $D$ is  the spatially-independent
``growth factor'' in the linear evolution of density perturbations and
$f \equiv \frac{d\ln{D}}{d\ln{a}}$ is the linear growth rate where $a$ is the scale factor  \cite{2006PhRvD..73l3526H,2011ApJ...741...67D}.

\begin{figure}[h]
\centering
\includegraphics[width=0.44\textwidth]{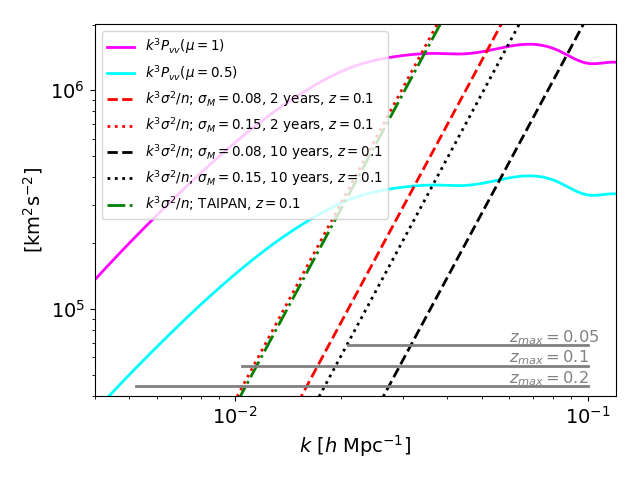}
\caption{Volume-weighted peculiar velocity power spectrum $k^3P_{vv}(z=0)$ for $\mu \equiv \cos{(\hat{k} \cdot \hat{r})}=1, 0.5$ (magenta, cyan) 
where $\hat{r}$ is the line of sight, as predicted for General Relativity in the linear regime.
Overplotted are peculiar-velocity power-spectrum shot noise  (diagonal lines) for various observing parameters.  Red shows the shot noise expected from a 2-year LSST survey
while black shows a 10-year LSST survey.  The dotted and dashed lines indicate the assumed intrinsic magnitude dispersion, using 0.08 (dashed) or 0.15 mag (dotted).  The expected shot
noise from TAIPAN is shown in green (dash-dotted). 
The bottom solid grey horizontal lines show the approximate range of $k$ expected to be used in surveys with corresponding
redshift depths $z_{\rm max}$.
\label{power:fig}}
\end{figure}

The $\Lambda$CDM prediction for the $z=0$ peculiar velocity power spectrum is shown in Figure~\ref{power:fig}. The growth of structure depends on gravity;
\cite{2007APh....28..481L} find that General Relativity, $f(R)$,  and DGP gravity follow the relation
$f \approx \Omega_M^\gamma$ with $\gamma=0.55, 0.42, 0.68$ respectively (see \cite{HUTERER201523} for a review
or these  models).  Using this parameterization to model gravity, peculiar velocity
surveys probe $\gamma$ through $fD$, whose $\gamma$-dependence is plotted 
in Figure~2 of  \cite{1475-7516-2013-04-031}.

Peculiar velocity surveys have already been  used to measure $fD$ (also referred to as $f\sigma_8$), though not to a level where  gravity models can be precisely distinguished.
 \cite{2017MNRAS.471..839A} use 6dFGS peculiar velocities using  Fundamental Plane distances of elliptical galaxies to estimate absolute magnitudes
 with
 $\sim 0.43$~mag  precision, yielding a 15\% uncertainty in $fD$ at $z\approx 0$.
The upcoming 
TAIPAN survey \cite{2017PASA...34...47D} will obtain Fundamental Plane galaxies with densities of $n_g \sim 10^{-3}h^3$\,Mpc$^{-3}$,
and the WALLABY+WNSHS surveys \cite{2008ExA....22..151J} will obtain Tully-Fisher distances (based on the $\sim 0.48$~mag calibration of absolute magnitude based on the  HI 21cm line width)
of galaxies with densities $n_g \sim 2\times 10^{-2} - 10^{-4} h^3$\,Mpc$^{-3}$ from
$z=0-0.1$ covering 75\% of the sky.
These surveys combined are projected to have 3\% uncertainties in $fD$ \cite{2017MNRAS.464.2517H}.
For reference, DESI projects a 10\% precision of $fD$ at $z \approx 0.3$  by looking 
for signatures (Redshift Space Distortions; RSD) expected from galaxies infalling toward mass overdensities.
Relative to galaxies with  Fundamental Plane or Tully-Fisher distances, 
SN~Ia host galaxies currently have significantly lower number density but have better per-object peculiar velocity precision.
Existing SN~Ia samples
have been used to test and ultimately find spatial correlations in peculiar velocities that may be attributed to the growth of structure
\cite{PhysRevLett.99.081301,2008MNRAS.389L..47A,2014MNRAS.444.3926J,2015JCAP...12..033H, 2017JCAP...05..015H}.
SNe~Ia discovered by ASAS-SN, ATLAS, and ZTF \cite{2014ApJ...788...48S,2018PASP..130f4505T,2019PASP..131a8002B} over the next several years will provide first probative measures of $fD$ at $z<0.1$.

Two advances in the upcoming decade will make SN~Ia peculiar velocities more powerful.
First, the precision of SN~Ia distances can be improved.  The commonly-used empirical 2-parameter spectral model yields  absolute magnitude
dispersion $\sigma_M \gtrsim 0.12$~mag.  However, SNe transmit more information than just the light-curve shape and single color used in current SN models.
Recent studies indicate that with the right data, SN absolute
magnitudes can be calibrated to $\sigma_M \lesssim 0.08$ mag \cite[see e.g.][]{2012MNRAS.425.1007B, 2015ApJ...815...58F}. 
Though not yet
established, it is anticipated that such a reduction in intrinsic dispersion comes with a reduction in the magnitude bias correlated with host-galaxy properties
that is observed using current calibrations.  At this precision the intrinsic velocity dispersion  at $z=0.028$ is  $300$~km\,s$^{-1}$, i.e.\ a single SN~Ia  is of such quality as to
measure a peculiar velocity with $S/N \sim 1$.
 If corrections of all SNe~Ia are not possible, the use of SN~Ia subclasses is an option though at the expense of reducing the
numbers of velocity probes.
Secondly,  in the upcoming decade cadenced wide-field imaging surveys such as ZTF and LSST
  will increase the number of identified  $z<0.3$ Type~Ia supernovae from the hundreds to the
hundreds of thousands; over the course of 10-years, LSST will find $\sim150,000$ $z<0.2$,  $\sim520,000$ $z<0.3$ SNe~Ia
 for which good light curves can be measured, corresponding to a  number density of $n \sim 5\times 10^{-4}h^3$\,Mpc$^{-3}$.
  This sample has comparable
 number density and more galaxies at deeper redshifts than projected by WALLABY and TAIPAN.  With similar densities,
 the (two) ten-year SN~Ia survey will have
 a (6) 29$\times$ reduction in shot-noise, $\sigma^2_M/n$, relative to the Fundamental Plane survey of TAIPAN.

Given these  advances, supernovae discovered by wide-field searches in the next decade will 
 be able to tightly constrain the growth of structure in the low-redshift Universe.
For example, 
over the course of a decade a SN survey relying on LSST discoveries 
plus spectroscopic redshifts can produce  4--14\% uncertainties in $fD$ in $0.05$ redshift bins from $z=0$ to 0.3, cumulatively giving
2.2\% uncertainty on $fD$  within this interval, where at
$0<z<0.2$ most of the probative power comes from peculiar velocities and at higher redshifts from RSD  \cite{2017ApJ...847..128H}.

\section{Testing Gravity with Peculiar Velocity Surveys}
While the growth rate $fD$ can be used to test several aspects of physics beyond the standard cosmological model (e.g.\ dark matter clustering, dark energy evolution), our scientific interest is in probing gravity,
so here we focus  on the growth index $\gamma$.
To illustrate the distinction,
 $\frac{d(\ln{fD})}{d\gamma} = \ln\Omega_M + \int \Omega_M^{\gamma} \ln{\Omega_M}\, d\ln{a} \approx -1.68, -0.75,-0.37$ at $z=0,0.5,1.0$ respectively
 in $\Lambda$CDM;  two surveys with the same fractional precision  in $fD$ will have different precision in $\gamma$,
 with the one at lower redshift providing the tighter constraint. 
 In this section, we demonstrate that peculiar velocity surveys in the upcoming decade can measure $\gamma$ precisely
for a range of survey-parameter choices.
 
We project uncertainties  on the growth index, $\sigma_\gamma$ for a suite of idealized surveys
using a Fisher matrix analysis similar to that of \cite{2017ApJ...847..128H, 2017MNRAS.464.2517H}
(there is an alternative approach using an estimator for the mean pairwise velocity
\cite{2011PhRvD..83d3004B}).
The ``cross-correlation'' analysis incorporates both galaxy overdensities and peculiar velocities.
The Fisher information matrix is
\begin{align}
F_{ij} 
& = \frac{\Omega}{8\pi^2} \int_{r_{\rm min}}^{r_{\rm max}}  \int_{k_{\rm min}}^{k_{\rm max}}  \int_{-1}^{1} r^2 k^2 \text{Tr}\left[ C^{-1} \frac{\partial C}{\partial \lambda_i} C^{-1}
\frac{\partial C}{\partial \lambda_j} \right] d\mu\,dk\,dr
\label{fisher:eqn}
\end{align}
where
\begin{equation}
C(k,\mu)  =
  \begin{bmatrix}
   P_{\delta \delta}(k,\mu) + \frac{1}{n} &
   P_{v\delta}(k,\mu)  \\
   P_{v\delta}(k,\mu)  &
  P_{vv}(k,\mu) + \frac{\sigma^2}{n}
   \end{bmatrix}
\label{cov:eq}
\end{equation}
and the parameters considered are $\lambda \in \{\gamma,bD, \Omega_{M_0}\}$.
The parameter dependence enters through $fD$ in the relations $P_{vv}\propto (fD\mu)^2$, the SN~Ia host-galaxy count overdensity
power spectrum $P_{\delta \delta }\propto (bD + fD\mu^2)^2$, and the galaxy-velocity cross-correlation $P_{vg}
\propto  (bD + fD\mu^2)fD$, where $b$ is the galaxy bias and $\mu\equiv \cos{(\hat{k} \cdot \hat{r})}$ where $\hat{r}$ is the direction of
the line of sight.  While the $bD$ term does contain information on $\gamma$, its constraining power is not used here.
Both $f$ and $D$ depend on $\Omega_M=\frac{\Omega_{M_0}}{\Omega_{M_0} + (1-\Omega_{M_0})a^3}$.
The uncertainty in $\gamma$ is $\sigma_\gamma = \sqrt{\left(F^{-1}\right)_{\gamma\gamma}}$.
Non-GR models may  also predict a change in the scale-dependence of the growth or non-constant $\gamma$, such observations provide additional leverage in probing gravity
but are not considered here.

The uncertainty  $\sigma_\gamma$ of a survey depends on its solid angle $\Omega$, depth
given by the comoving distance out to the maximum redshift $r_{\rm max}=r(z_{\rm max})$, duration
$t$ through $n=\epsilon \phi t$ where $\phi$ is the observer-frame SN~Ia rate
and $\epsilon$ is the sample-selection efficiency, and
the intrinsic SN~Ia magnitude dispersion 
through the resulting peculiar velocity intrinsic dispersion $\sigma \approx (\frac{5}{\ln{10}} \frac{1+z}{z})^{-1} \sigma_M$.

We consider SN peculiar velocity surveys for a range of redshift depths $z_{\rm max}$ for durations of $t=2$ and 10~years.
The other survey parameters
$\Omega=3\pi$, $\epsilon=0.65$, $\sigma_M=0.08$~mag are fixed.
The $k$-limits are  taken to be $k_{\rm min} = \pi/r_{\rm max}$  and 
$k_{\rm max} = 0.1$~$h$\,Mpc$^{-1}$.
A minimum distance $r_{\rm min}=r(z=0.01)$ is imposed as our analysis assumes that peculiar velocities are significantly  smaller than
the cosmological redshift.
The sample-selection efficiency $\epsilon$ is redshift-independent, i.e.\ the native redshift distribution is not sculpted.
The input bias of SN~Ia host galaxies is set as $b=1.2$.  An independent measurement of $\Omega_{M_0}=0.3\pm0.005$
is included and is a non-trivial contributor to the $\gamma$ constraint.  Number densities are taken to be
direction-independent, neglecting the slight declination-dependence of SN-survey time windows

\begin{figure}
\centering
\includegraphics[width=0.44\textwidth]{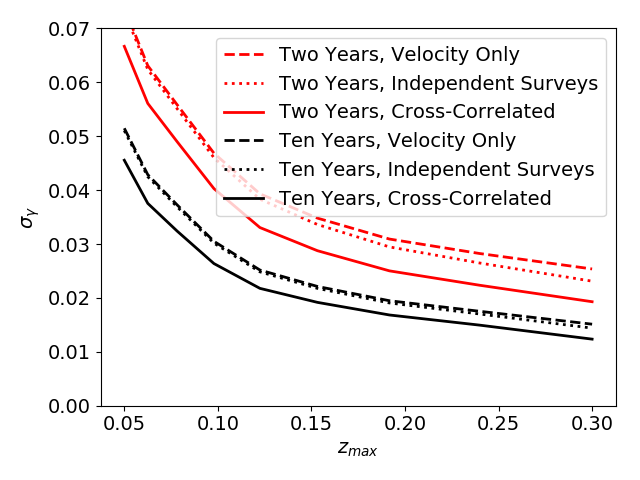}
\caption{The projected uncertainty in $\gamma$, $\sigma_\gamma$, achieved by two-year (red) and ten-year (black) SN~Ia surveys of varying depth $z_{\rm max}$.
For each survey uncertainties are based on three types of analyses:  using only peculiar velocities (dashed); using
both RSD and peculiar velocities independently (dotted);  using
both RSD, peculiar velocities, and their cross-correlation (solid).
\label{var:fig}}
\end{figure}

All the surveys considered  provide meaningful tests of gravity.
The projected uncertainty in $\gamma$ achieved by the suite of surveys are shown in Figure~\ref{var:fig}.
The primary result is for the cross-correlation analysis that uses overdensities (RSD), peculiar velocities, and their cross-correlations.
The short and shallow,  2-year, $z_{\rm max}=0.11$ survey has $\sigma_\gamma \sim 0.038$,
which can distinguish  between  General Relativity, $f(R)$,  and DGP gravities at the $>3\sigma$ level.
The 10-year survey performance asymptotes at $z_{\rm max} \sim0.2$ at a precision of  $\sigma_\gamma \sim 0.01$.
Figure~\ref{var:fig} also shows uncertainties based on two other analyses, one that only uses peculiar velocities, and
one that combines independent RSD and peculiar velocity results.
Peculiar velocities alone
account for much of the probative power of the surveys. RSD alone do not provide significant constraints.
However, considering RSD and velocity cross-correlations decreases $\sigma_\gamma$ by  $\sim 20$\%.
The implication is that there are important
$k$-modes that are  sample variance limited either in  overdensity and/or peculiar velocity who benefit from the
sample-noise suppression engendered by cross-correlations.

\begin{figure}
\centering
\includegraphics[width=0.44\textwidth]{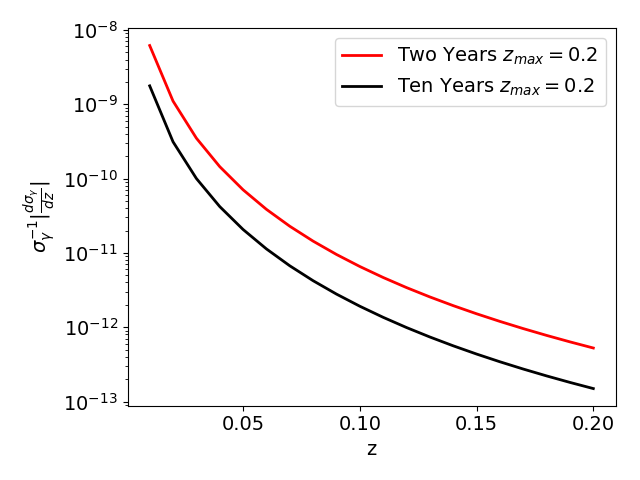}
\includegraphics[width=0.44\textwidth]{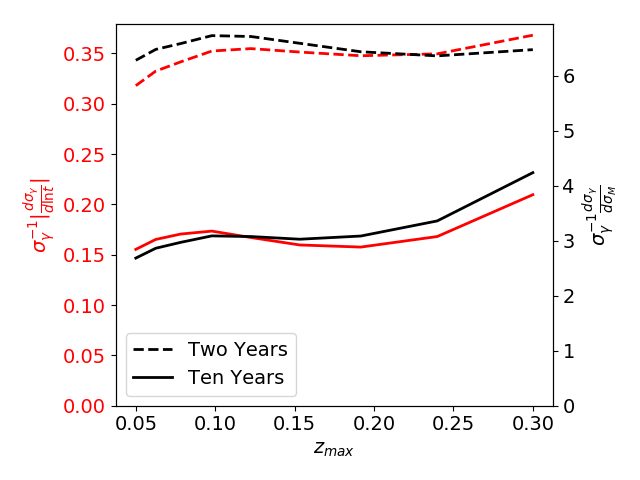}
\caption{Left: $\sigma_\gamma^{-1}|\partial \sigma_\gamma / \partial z|$  after two and ten years for a survey with limiting  depth $z_{\rm max}=0.2$. 
Right: $\sigma_\gamma^{-1}|\partial \sigma_\gamma / \partial \ln{t}|$ (red) and $\sigma_\gamma^{-1}\partial \sigma_\gamma / \partial \sigma_M$ (black) each as a function of
 $z_{\rm max}$ for two- (dashed) and ten-year (solid) surveys.
\label{dvar:fig}}
\end{figure}

Survey performance is examined in more detail by considering how $\sigma_\gamma$ in the cross-correlation analysis
changes with respect to the 
survey parameters  $\Omega$, $z_{\rm max}$, $t$, and $\sigma_M$, and also with respect to differential redshift bins
within a given survey.  Though not directly a survey parameter, we also examine changes with respect to our fiducial choice of  $k_{\rm max}$. \newline
{\it Solid Angle $\Omega$:}
The Fisher Matrix $F$ is  proportional to the survey solid angle $\Omega$ so $\sigma_\gamma \propto \Omega^{-1/2}$.
{\it Differential Redshift Bin $z$:}  Certain redshifts constrain $\gamma$ more strongly than others.  If at a given moment of a survey we had
a set of SNe~Ia from which to choose, it turns out the one with the lowest redshift would be preferred.
This is demonstrated to be the case at the end of both 2- and 10-year surveys with $z_{\rm max}=0.2$.
The left panel of Figure~\ref{dvar:fig} shows
 $|\partial \sigma_\gamma / \partial z|$, which
for both surveys monotonically decreases from $z=0.01$ out to $z=0.2$.  If we had
to sculpt the distribution, the preference would be to cut out the highest redshift bins
resulting in a decreased $z_{\rm max}$.   The optimal redshift distribution is thus the unsculpted SN-discovery distribution truncated by $z_{\rm max}$. 
\newline
{\it Redshift Depth $z_{\rm max}$:}
Increasing the survey redshift depth increases the $\gamma$ precision.  The differential improvement in $\sigma_\gamma$ plateaus at
$z_{\rm max} \sim 0.2$ as seen in Figure~\ref{var:fig}. \newline
{\it Survey duration $t$; Intrinsic Magnitude Dispersion $\sigma_M$:}
An increased survey duration accumulates more supernovae, decreasing shot noise and  increasing the 
precision in $\gamma$ for all the surveys considered.
The surveys we consider have varying relative contributions of sample variance and shot noise: those that have a larger shot-noise contribution (i.e.\ 
shorter surveys and those with higher $z_{\rm max}$) benefit more from extending the survey duration.
Like survey duration, intrinsic magnitude dispersion is related to survey performance through the shot noise and thus
has a similar relationship with $\sigma_\gamma$; the effect of duration and magnitude dispersion are shown in the right-panel plot of Figure~\ref{dvar:fig} as  $\sigma_\gamma^{-1}|\partial \sigma_\gamma / \partial \ln{t}|$ 
and $\sigma_\gamma^{-1}\partial \sigma_\gamma/\partial \sigma_M$ 
as a function of
 $z_{\rm max}$ for two- and ten-year surveys.\newline
{\it Minimum length scale, maximum wavenumber $k_{\rm max}$:} There is a minimum length scale at which density and velocity distributions are reliably predicted from theory.  Changes in this scale engender  fractional changes in the $\gamma$ precision as  $\sigma_\gamma^{-1}\partial \sigma_\gamma/\partial k_{\rm max} = 0.0050$ at $k_{\rm max}=0.1h$\,Mpc$^{-1}$,
 which  is survey-independent.

\section{Conclusions}

In the next decade,
the high number of SN discoveries together with improved precision in their distance precisions will make $z<0.3$ SNe~Ia, more so than
galaxies,  powerful probes of gravity through their effect  on the growth of structure.  Different survey strategies can be adopted to take advantage of these
supernovae, and in this White Paper we present a formalism and code (available
at \url{http:tiny.cc/PVScience})
by which their scientific merits can be assessed and present results for a range of options.

No other probe of growth of structure or tracer of peculiar velocity can alone provide comparable precision on  $\gamma$ in the next decade.
At low redshift, the RSD measurement is quickly sample variance limited (as are the planned DESI BGS and 4MOST surveys) making peculiar velocities the only 
precision probe of $fD$.
TAIPAN and a TAIPAN-like DESI BGS will be able to measure FP distances for nearly all usable nearby galaxies, so at low-$z$ the Fundamental Plane peculiar-velocity
technique will  saturate at a level that is not competitive with a  2-year SN survey.

Combined low-redshift peculiar velocity and high-redshift RSD $fD$ measurements are highly complementary as together they probe the
$\gamma$-dependent shape of $fD(z)$ (not just its normalization) and potential scale-dependent influence of gravitational models, since low-
and high-redshift surveys are weighted by lower and higher $k$-modes respectively.
SN~Ia peculiar velocity surveys are of the highest scientific
interest and we encourage
the community to develop aggressive surveys in the pursuit of testing  General Relativity and probing gravity. 

\bibliographystyle{plain}
\bibliography{/Users/akim/Documents/alex}

\end{document}